\documentclass{article}

\usepackage{graphicx}
\usepackage{amsmath}

\title{Detecting a trend change in cross-border epidemic transmission}

\author{Yoshiharu Maeno \\ Social Design Group \\ email: maeno.yoshiharu@socialdesigngroup.com}

\begin{document}

\maketitle

\begin{abstract}
A method for detecting a trend change in cross-border epidemic transmission is developed for a standard epidemiological SIR compartment model and a meta-population network model. The method is applicable to investigating the efficacy of the implemented public health intervention in managing infectious travelers across borders from a time series of the number of new cases reported in multiple geographical regions. It is found that the change point of the probability of travel movements was one week after the WHO worldwide alert on the SARS outbreak in 2003. The alert was effective in managing infectious travelers. On the other hand, it is found that the probability of travel movements did not change at all for the flu pandemic in 2009. The pandemic did not affect potential travelers despite the WHO alert.
\end{abstract}

\section{Introduction}

An infectious disease outbreak is a complex stochastic phenomenon in a spatially heterogeneous medium\cite{New02}, \cite{Bog02}. The analysis of the observations on an outbreak includes many tasks, which range from reproducing the growing number of cases at an infected city to detecting the omen and predicting the onset of an outbreak at neighboring uninfected cities. Among them, detecting a trend change in cross-border epidemic transmission is a task of particular interest to public health authorities.

When the public health authorities issue an alert on the risk of massive community transmission commencing worldwide, public awareness may threaten potential travelers into refraining from travels. Then it becomes less probable that infectious travelers cross national or regional borders. Border health screening at airports and isolation of infectious travelers work similarly to these voluntary risk-averse behaviors. The community transmission decelerates if such a social distancing works effectively in controlling cross-border exposure. This is an example of a beneficial trend change in epidemic transmission. Detecting whether a trend changes or not helps the public health authorities confirm the efficacy of the current practices of public health intervention and design a more effective public health program.

Early detection helps the public health authorities eliminate the bottleneck of the public health intervention and contain the epidemic quickly. Some previous related works apply a model selection technique to detecting the change of transmission parameters. Other studies apply regression analysis to the early detection of the onset of an outbreak. The dataset studied by most of these works is a uni-variate time series for the number of cases in a single geographic region. These methods are not applicable to the investigation of the travel movements between geographic regions and the cross-border epicemic transmission by infectious travelers. Few studies address a correlated multivariate time series in multiple geographic regions.

In this study, a method for detecting a trend change in cross-border epidemic transmission is developed for a standard epidemiological SIR compartment model and a meta-population network model. The method is applicable to investigating the efficacy of the implemented public health intervention in managing infectious travelers across borders from a time series of the number of new cases reported in multiple geographical regions.

\section{Related works}

{\bf Studies in theoretical, experimental, and applied physics devote much effort to developing Bayesian statistical methods to solve an inverse problem for a dynamic system\cite{Tou11}. An inverse problem  of interest to physics ranges from determining the value of an endogenous parameter, revealing an  unknown boundary condition, and finding the initial condition of a variable to detecting the trend  change in an exogenous parameter. The difficulty in solving an inverse problem is the complexity in  computing a posterior probability density function and a Bayes factor. A posterior probability  density function is computed with a belief propagation algorithm\cite{Alt14}. A marginal probability  density function is computed by factorizing the joint posterior probability density function into the  family of parameterized normal distributions\cite{Sch11}. A series of Bayes factors are computed  analytically\cite{Ens10}. Those studies present that Bayesian statistical methods contribute to significant findings on the nature of a dynamic system.}

A trend change in a reproductive ratio of a H5N1 avian influenza is investigated with an anomaly detection technique, and a big increase in June 2006 is found\cite{Bet08}. The trend change in the parameters for infection and recovery is analyzed in the early, middle, and late phases of a SARS epidemic in 2003 with an approximate Bayesian computation technique\cite{Wal10}. A degree distribution is selected from representative degree distributions of a heterogeneous contact network between hosts for measles, gonorrhea, and norovirus outbreaks\cite{Sta13}. A compartment model with stratification by age and cross immunity for multiple strains fits the number of seasonal influenza cases the best\cite{Tru12}. A sequential importance sampling technique is applied in selecting a compartment model from candidate compartment models with a sub-divided compartments, an additional transition between compartments, or a time delay of transitions for a common cold outbreak\cite{Ton09}. In those model selection techniques\cite{Tou11}, a Bayes factor or a Schwarz's Bayesian information criterion\cite{Sch11}, \cite{Ens10}, \cite{Bro03} indicate the relative goodness of fit between models.

Regression techniques and control charts are often applied to the early detection of an outbreak\cite{Unk12}. The change in an emergency department visit rate is analyzed for a real time syndromic surveillance\cite{Rei03}. An autoregressive integrated moving average model is developed for the early detection of such a bioterrorist attack as an anthrax release and contamination.  A physician visit rate is predicted from the fraction of the queries to an online search engine\cite{Gin09}. The increase in this rate signals an impending outbreak. The current level of disease activity is modeled as a hidden state variable in a Markov switching model\cite{Lu10}. This model is applied to the prospective detection of cryptosporidiosis and anthrax outbreaks. A multivariate regression method is developed to detect an outbreak more robustly from the difference between the observations in multiple geographical regions\cite{Sch12}.

\section{Problem}

The entire population is sub-divided into distinct sub-populations in multiple geographical regions. The geographical regions are nodes $n_{i} \ (i=0,1,\dots,N-1)$. $N$ is the number of nodes. The transportation between geographical regions is a pair of unidirectional links between nodes. Observations are made at times $t_{d} \ (d=0,1,\dots,D-1)$ at every node. $D$ is the number of observations. The time interval between observations is $\Delta t =t_{d+1} - t_{d}$. The cumulative number of new cases until $t_{d}$ is a vector variable $\mbox{\boldmath{$J$}}(t_{d})=(J_{0}(t_{d}),J_{1}(t_{d}),\dots,J_{N-1}(t_{d}))$ where the elemets $J_{i}(t)$ are the cumulative number at the node $n_{i}$. The number of new cases between subsequent observations is $\Delta \mbox{\boldmath{$J$}}(t_{d}) = \mbox{\boldmath{$J$}}(t_{d+1}) - \mbox{\boldmath{$J$}}(t_{d})$. The time sequence $\mbox{\boldmath{$D$}} = \{ \Delta \mbox{\boldmath{$J$}}(t_{d}) \} \ (d=0,1,\dots,D-1)$ forms a dataset. An example of a dataset is a bundle of the daily reports on new cases from hospitals in neighboring cities.

The problem is to detect from $\mbox{\boldmath{$D$}}$ whether two probability parameters change or not, which govern cross-border epidemic transmission. The cross-border epidemic transmission ensues from the travel of an infectious person at an infected node to an uninfected neighboring node, and local transmission from the traveler to susceptible residents there. The probability of a person moving between nodes per a unit time is the product of two factors. One is the coefficient of proportionality $\gamma$, which determines the absolute trend of movements. The other is the relative volume of spatially heterogeneous movements between nodes. The relative volume can be determined by an empirical law as a function of the topology of links. The value of $\gamma$ may change once at an unknown change point $t_{{\rm c}}^{[\gamma]}$. Neither the topology of links nor the relative volume of movements change in this study. The parameter $\alpha$ is the probability of an infectious person contacting a person and infecting the person per a unit time. It governs a local outbreak. The value of $\alpha$ may also change once at an unknown change point $t_{{\rm c}}^{[\alpha]}$.

\section{Method}

\subsection{Time evolution}
\label{Time}

The mathematical model of cross-border transmission is a special case of a stochastic reaction-diffusion process, which is the integration of a standard epidemiological SIR compartment model and a meta-population network model\cite{Bar08}. 

The presence of links in the meta-population network is represented by an adjacency matrix $\mbox{\boldmath{$l$}}$. If a pair of uni-directional links is present between $n_{i}$ and $n_{j}$, $l_{ij}$ and $l_{ji}$ are 1, and 0 otherwise. The probability of a person moving from the $i$-th node to the $j$-th node is $\gamma_{ij}$, which forms a $N \times N$ matrix. The value of its elements is derived by an empirical law\cite{Mae10}, \cite{Bar04} in eq.(\ref{gammacalc}). The nodal degree of the $i$-th node is $k_{i} = \sum_{j=0}^{N-1} l_{ij}$. The empirical law is valid generally for the world-wide airline transportation network\cite{Col06}.
\begin{eqnarray}
\gamma_{ij}= \frac{l_{ij} \sqrt{k_{i} k_{j}}}{\sum_{j=0}^{N-1} l_{ij} \sqrt{k_{i} k_{j}}} \gamma.
\label{gammacalc}
\end{eqnarray}

The state of a person changes from a susceptible state, through an infectious state, to a removed (recovered) state. The time dependent variables $S_{i}(t)$, $I_{i}(t)$, $R_{i}(t)$, and $J_{i}(t)$ is the number of susceptible persons, infectious persons, removed persons, and the cumulative number of new cases at the $i$-th node at $t$. The parameter $\beta$ is the probability of an infectious person removed per a unit time. In this study, $\beta$ does not change. The reproductive ratio is given by $R=\alpha/\beta$. Movements, infection, and removal are Markovian stochastic processes.

The time evolution of those variables is given by a Langevin equation\cite{Huf04}, which is a system of stochastic differential equations. The Langevin equation is given by eq.(\ref{apdI/dt}) and (\ref{dJ/dt}) in the early phase of the outbreak when $I_{i} \ll S_{i}$ and $R_{i} \ll S_{i}$ hold true\cite{Mae10}. The statistical property of the terms $\xi(t)$ is the Gaussian white noise.
\begin{eqnarray}
\frac{{\rm d} I_{i}(t)}{{\rm d} t} &=& \alpha I_{i}(t) - \beta I_{i}(t) 
+ \sum_{j=0}^{N-1} \gamma_{ji} I_{j}(t) - \sum_{j=0}^{N-1} \gamma_{ij} I_{i}(t) \nonumber \\
&+& 
\sqrt{\alpha I_{i}(t)} \xi^{[\alpha]}_{i}(t) - \sqrt{\beta I_{i}(t)} \xi^{[\beta]}_{i}(t) \nonumber \\
&+& \sum_{j=0}^{N-1} \sqrt{\gamma_{ji} I_{j}(t)} \xi^{[\gamma]}_{ji}(t) - \sum_{j=0}^{N-1} \sqrt{\gamma_{ij} I_{i}(t)} \xi^{[\gamma]}_{ij}(t).
\label{apdI/dt}
\end{eqnarray}
\begin{eqnarray}
\frac{{\rm d} J_{i}(t)}{{\rm d} t} = \alpha I_{i}(t) + \sqrt{\alpha I_{i}(t)} \xi^{[\alpha]}_{i}(t).
\label{dJ/dt}
\end{eqnarray}

Equivalently, the time evolution of the joint probability density function of the corresponding probability variable vector is given by a Fokker-Planck equation, which is a partial differential equation. The Fokker-Planck equation is converted to a system of ordinary differential equations to calculate the moments of the probability variables one order after another\cite{Mae11}. The total cumulative number of new cases until $t$ is given by $J(t)= \sum_{i=0}^{N-1} J_{i}(t)$. The mean of $J$ at $t$ is given by eq.(\ref{mJsol}) when the value of $\alpha$ does not change. $I_{0}$ is the initial number of infectious persons.
\begin{eqnarray}
m^{{\rm [J]}}(t|I_{0},\alpha,\beta) =I_{0} (\frac{\alpha}{\alpha-\beta} \exp(\alpha-\beta)t - \frac{\beta}{\alpha-\beta}).
\label{mJsol} 
\end{eqnarray}

The variance about the mean of $J$ at $t$ is given by eq.(\ref{mJsol}).
\begin{eqnarray}
v^{{\rm [J]}}(t|I_{0},\alpha,\beta) &=& I_{0} [ \frac{\alpha^{2}(\alpha+\beta)}{(\alpha-\beta)^{3}} \exp 2(\alpha-\beta)t \nonumber \\
&-& \{\frac{\alpha(\alpha+\beta)}{(\alpha-\beta)^{2}} + \frac{4\alpha^{2} \beta}{(\alpha-\beta)^{2} }t\} \exp (\alpha-\beta)t - \frac{\alpha\beta(\alpha+\beta)}{(\alpha-\beta)^{3}} ].
\label{vJJsol}
\end{eqnarray}

The moments of $I_{i}$ at $t$ are not derived in closed forms. The mean of $I_{i}$ at $t+\Delta t$ for small $\Delta t$ is given by eq.(\ref{m(deltat)}) when the values of $\alpha$ and $\gamma$ do not change. The coefficients $a_{ip}$ are defined by eq.(\ref{aip}).
\begin{eqnarray}
m^{{\rm [I]}}_{i}(t+\Delta t|\alpha,\beta,\mbox{\boldmath{$\gamma$}}) = I_{i}(t) + \sum_{p} a_{ip} I_{p}(t) \Delta t +O(\Delta t^{2}).
\label{m(deltat)}
\end{eqnarray}
\begin{eqnarray}
a_{ip} = (\alpha-\beta-\sum_{j'} \gamma_{ij'})\delta_{ip} + \gamma_{pi}.
\label{aip}
\end{eqnarray}

The covariance about the mean between $I_{i}$ and $I_{j}$ at $t+\Delta t$ for small $\Delta t$ is given by eq.(\ref{v(deltat)}). The coefficients $b_{ijp}$ are defined by eq.(\ref{bijp}) where $\delta$ is the Kronecker's symbol.
\begin{eqnarray}
v^{{\rm [I]}}_{ij}(t+\Delta t|\alpha,\beta,\mbox{\boldmath{$\gamma$}}) = \sum_{p} b_{ijp} I_{p}(t) \Delta t +O(\Delta t^{2}).
\label{v(deltat)}
\end{eqnarray}
\begin{eqnarray}
b_{ijp} = \{(\alpha+\beta+\sum_{j'} \gamma_{ij'}) \delta_{ip}+\gamma_{pi} \} \delta_{ij} - \gamma_{ij} \delta_{ip} - \gamma_{ji} \delta_{jp}.
\label{bijp}
\end{eqnarray}

If $\gamma$ changes from $\gamma_{1}$ to $\gamma_{2}$ at time $t=t_{{\rm c}}^{{[\gamma]}}$, or $\alpha$ changes from $\alpha_{1}$ to $\alpha_{2}$ at $t=t_{{\rm c}}^{{[\alpha]}}$, the moments satisfy the boundary conditions at the change points. Their formulae become more complicated than eq.(\ref{mJsol}) through (\ref{bijp}). The skewness, kurtosis, and higher order moments are ignored in this study. The probability density function $P(J,t)$ is approximated as a normal distribution with the mean in eq.(\ref{mJsol}) and the variance in eq.(\ref{vJJsol}). The joint probability density function $P(\mbox{\boldmath{$I$}},t)$ is approximated as a multivariate normal distribution with the mean and covariance in eq.(\ref{m(deltat)}) through (\ref{bijp}).

\subsection{Problem decomposition}
\label{Decomposition}

A trend change is detected with a model selection technique\cite{Has09}. Model selection is the task of selecting a model which fits a dataset the best from a set of candidate models. A model without any change points of parameters is compared with a model with a change point. The model which fits a given dataset better is selected. The latter model is selected when the trend changes while the former model is selected when the trend does not change. Two computationally efficient model selectors are presented in \ref{ModelSelector}.

The problem is decomposed into a sequence of two sub-problems. The first preparatory sub-problem is the $\alpha$ problem to detect the change in $\alpha$ from the time sequence of $J(t_{d})$. The solution of the $\alpha$ problem does not depend on the value of $\mbox{\boldmath{$\gamma$}}$ because the time evolution of $J(t)$ is determined merely by the value of $\alpha$ and $\beta$. The value of $\beta$ does not change in this study. The model selectors are applied here. If the model without any change points is selected, the estimates of $\alpha$ and $\beta$ are obtained with a maximal likelihood estimation or a maximal a posteriori estimation. The estimates are represented by $\hat{\alpha}$ and $\hat{\beta}$. Similarly, if the model with a change point is selected, $\hat{\alpha}_{1}$, $\hat{\alpha}_{2}$, $\hat{t}_{{\rm c}}^{{[\alpha]}}$, and $\hat{\beta}$ are obtained. The value of $\alpha$ changes from $\hat{\alpha}_{1}$ to $\hat{\alpha}_{2}$ at time $t=\hat{t}_{{\rm c}}^{{[\alpha]}}$.

The second sub-problem is the $\gamma$ problem to detect the change in $\gamma$ from the time sequence of $\mbox{\boldmath{$I$}}(t_{d})$. As a preparation to solve the $\gamma$ problem, given either $(\hat{\alpha}, \hat{\beta})$ or $(\hat{\alpha}_{1}, \hat{\alpha}_{2}, \hat{t}_{{\rm c}}^{{[\alpha]}}, \hat{\beta})$, the value of the elements of $\mbox{\boldmath{$l$}}$ is obtained from the time sequence of $\mbox{\boldmath{$I$}}(t_{d}) \approx \Delta \mbox{\boldmath{$J$}}(t_{d}) / \hat{\alpha} \Delta t$ with a maximal likelihood estimation\cite{Mae10}. It is represented by $\hat{\mbox{\boldmath{$l$}}}$. The adjacency matrix does not change. The model selectors are applied here. If the model without any change points is selected, the value of $\hat{\gamma}$ is obtained. If the model with a change point is selected, the value of $\hat{\gamma}_{1}$, $\hat{\gamma}_{2}$, and $\hat{t}_{{\rm c}}^{{[\gamma]}}$ are obtained. The value of $\gamma$ changes from $\hat{\gamma}_{1}$ to $\hat{\gamma}_{2}$ at time $t=\hat{t}_{{\rm c}}^{{[\gamma]}}$.

\subsection{Model selector}
\label{ModelSelector}

The relative goodness of fit between two candidate models is given by a Bayes factor\cite{Kas95} in Bayesian statistics. The definition of the Bayes factor $F$ is the ratio of two posterior probabilities in eq.(\ref{genBF}) when one model is parameterized by a vector quantity $\mbox{\boldmath{$\theta$}}_{1}$ and the other model by $\mbox{\boldmath{$\theta$}}_{2}$.
\begin{eqnarray}
F = \frac{
\int_{\mbox{\boldmath{$\theta$}}_{1} \in \mbox{\boldmath{$\Theta$}}_{1}} L_{1}(\mbox{\boldmath{$\theta$}}_{1}) P_{1}(\mbox{\boldmath{$\theta$}}_{1}) {\rm d}\mbox{\boldmath{$\theta$}}_{1}
}{
\int_{\mbox{\boldmath{$\theta$}}_{2} \in \mbox{\boldmath{$\Theta$}}_{2}} L_{2}(\mbox{\boldmath{$\theta$}}_{2}) P_{2}(\mbox{\boldmath{$\theta$}}_{2}) {\rm d}\mbox{\boldmath{$\theta$}}_{2}
}.
\label{genBF}
\end{eqnarray}

The likelihood function $L(\mbox{\boldmath{$\theta$}})$ equals to the probability density function $P(\mbox{\boldmath{$D$}}|\mbox{\boldmath{$\theta$}})$. P(\mbox{\boldmath{$\theta$}}) is the prior probability density function of the parameter vector. $\mbox{\boldmath{$\Theta$}}$ is the domain of definition for the parameter vector. If $F>1$, the first model fits the dataset better than the second model. The commonly used scale for interpretation is as follows\cite{Jef98}. If $10>F>3$, the selection of the first model is substantial. If $30>F>10$, the selection is strong. If $100>F>30$, the selection is very strong. If $F>100$, the selection is decisive. This interpretation applies to any pairs of models.

The likelihood functions to solve the $\alpha$ problem are given by eq.(\ref{bfsalpha1}) and eq.(\ref{bfsalpha2}). It is assumed that $P(\mbox{\boldmath{$D$}}|\mbox{\boldmath{$\theta$}})$ is a normal distribution with the calculated mean $m^{{\rm [J]}}$ and variance $v^{{\rm [J]}}$. Note that $I_{0}$ is also a parameter.
\begin{eqnarray}
L_{1}(\mbox{\boldmath{$\theta$}}_{1}) = \prod_{d=1}^{D-1} P_{{\rm N}}(J(t_{d})|I_{0},\alpha,\beta).
\label{bfsalpha1}
\end{eqnarray}
\begin{eqnarray}
L_{2}(\mbox{\boldmath{$\theta$}}_{2}) = \prod_{t_{d} \leq t_{{\rm c}}^{[\alpha]}} P_{{\rm N}}(J(t_{d})|I_{0},\alpha_{1},\beta)
\prod_{t_{d} > t_{{\rm c}}^{[\alpha]}} P_{{\rm N}}(J(t_{d})|I_{0},\alpha_{2},\beta).
\label{bfsalpha2}
\end{eqnarray}

The likelihood functions to solve the $\gamma$ problem are given by eq.(\ref{bfsalpha1}) and eq.(\ref{bfsalpha2}). It is that assumed that $P(\mbox{\boldmath{$D$}}|\mbox{\boldmath{$\theta$}})$ is a multivariate normal distribution with the calculated mean $\mbox{\boldmath{$m$}}^{{\rm [I]}}$ and variance $\mbox{\boldmath{$v$}}^{{\rm [I]}}$.
\begin{eqnarray}
L_{1}(\mbox{\boldmath{$\theta$}}_{1}) = \prod_{d=1}^{D-1} P_{{\rm N}}(\mbox{\boldmath{$I$}}(t_{d})|\gamma).
\label{bfsgamma1}
\end{eqnarray}
\begin{eqnarray}
L_{2}(\mbox{\boldmath{$\theta$}}_{2}) = \prod_{t_{d} \leq t_{{\rm c}}^{[\gamma]}} P_{{\rm N}}(\mbox{\boldmath{$I$}}(t_{d})|\gamma_{1})
\prod_{t_{d} > t_{{\rm c}}^{[\gamma]}} P_{{\rm N}}(\mbox{\boldmath{$I$}}(t_{d})|\gamma_{2}).
\label{bfsgamma2}
\end{eqnarray}

The value of the integrals in eq.(\ref{genBF}) is obtained for the likelihood functions in eq.(\ref{bfsalpha1}) through eq.(\ref{bfsgamma2}) neither analytically nor computationally efficiently. Two computationally efficient model selectors are presented in \ref{MrLS} and \ref{MLS}. One is a marginalized likelihood selector which calculates $F$ numerically with a Monte Carlo integration\cite{Rob10}. The other is a maximal likelihood selector which calculates a Schwarz's Bayesian information criterion\cite{Sch78} as a single point Gaussian approximation to obtain the value of $F$.

If the landscape of the likelihood function $L(\mbox{\boldmath{$\theta$}})$ has a single sharp peak at the maximal likelihood estimator $\hat{\mbox{\boldmath{$\theta$}}}$, that is the global maximum, the maximal likelihood selector tends to work more efficiently than the marginalized likelihood selector. The reason for this is that the single point Gaussian approximation is suitable for reproducing the peak while the Monte Carlo integration may be inaccurate if the density of random samples is too low to reproduce the peak. On the other hand, if the landscape is rugged with multiple peaks of similar altitude, or undulating gently in $\mbox{\boldmath{$\Theta$}}$, the maximal likelihood selector tends to be more erroneous than the marginalized likelihood selector. Which is more excellent depends on the nature of the $\alpha$ problem and $\beta$ problem, and the conditions like the dimension of the dataset $N$, the number of observations $D$, and the dimension of a parameter vector $|\mbox{\boldmath{$\theta$}}|$. The difference between the model selectors in detecting the trend change correctly is investigated in \ref{SynDatasets}.

\subsubsection{Marginalized likelihood selector}
\label{MrLS}

An approximate value of the integrals in eq.(\ref{genBF}) is obtained with a Monte Carlo integration in eq.(\ref{mci}). This is a technique for numerical integration with random numbers. Random samples $\{\mbox{\boldmath{$\theta$}}_{m}\} \ (m=0,1,\dots,M-1)$ are generated from the prior probability density function $P(\mbox{\boldmath{$\theta$}})$.
\begin{eqnarray}
\int_{\mbox{\boldmath{$\theta$}} \in \mbox{\boldmath{$\Theta$}}} L(\mbox{\boldmath{$\theta$}}) P(\mbox{\boldmath{$\theta$}}) {\rm d}\mbox{\boldmath{$\theta$}} \approx \frac{1}{M} \sum_{m=0}^{M-1} L(\mbox{\boldmath{$\theta$}}_{m}).
\label{mci}
\end{eqnarray}

Discriminating whether $F>1$ or $F<1$ from the approximate value by eq.(\ref{mci}) forms the marginalized likelihood selector.

\subsubsection{Maximal likelihood selector}
\label{MLS}

An approximate value of the integrals in eq.(\ref{genBF}) is obtained from the Bayesian information criterion $C$ in eq.(\ref{bic2}) by eq.(\ref{bic1}) . The formula for $C$ is derived by expanding $L(\mbox{\boldmath{$\theta$}})$ around the maximal likelihood estimator $\hat{\mbox{\boldmath{$\theta$}}}$ as a single point Gaussian approximation and by applying the Laplace's method\cite{Bis07} for calculating a finite integral. It is interpreted that $C$ is an absolute measure to quantify the best balance between the goodness of fit and model complexity. The model complexity is represented by $|\mbox{\boldmath{$\theta$}}|$. For example, $|\mbox{\boldmath{$\theta$}}|=5$ when $\mbox{\boldmath{$\theta$}}=(I_{0}, \alpha_{1}, \alpha_{2}, t_{{\rm c}}^{{[\alpha]}}, \beta)$.
\begin{eqnarray}
\int_{\mbox{\boldmath{$\theta$}} \in \mbox{\boldmath{$\Theta$}}} L(\mbox{\boldmath{$\theta$}}) P(\mbox{\boldmath{$\theta$}}) {\rm d}\mbox{\boldmath{$\theta$}} \approx \exp(-\frac{1}{2} C).
\label{bic1}
\end{eqnarray}
\begin{eqnarray}
C = - 2 \ln L(\hat{\mbox{\boldmath{$\theta$}}}) + |\mbox{\boldmath{$\theta$}}| \ln D.
\label{bic2}
\end{eqnarray}

Discriminating $F>1$ or $F<1$ from the approximate value by eq.(\ref{bic1}) forms the maximal likelihood selector.

\section{Result}

\subsection{Synthesized dataset}
\label{SynDatasets}

The model selectors are tested with synthesized datasets. The datasets are generated by solving the Langevin equation in eq.(\ref{dJ/dt}) numerically with a pseudo random number generator, and by recording the value of the variables at the times to make observations. The prior probability density function is uninformative in $\mbox{\boldmath{$\Theta$}}$. For the $\alpha$ problem,
\begin{itemize}
\item $P_{1}(I_{0},\alpha,\beta)$ for the model without any change points is a constant in $\mbox{\boldmath{$\Theta$}}_{1} = \{I_{0},\alpha,\beta |\ 0.9 J(t_{0}) \leq I_{0} \leq 1.1 J(t_{0})\ \land \ 0 \leq \alpha \leq 1\ \land \ 0 \leq \beta \leq 1 \}$
\item $P_{2}(I_{0}, \alpha_{1},\alpha_{2},t_{{\rm c}}^{[\alpha]},\beta)$ for the model with a change point is a constant in $\mbox{\boldmath{$\Theta$}}_{2} = \{I_{0}, \alpha_{1}, \alpha_{2}, t_{{\rm c}}^{[\alpha]}, \beta |\ 0.9J(t_{0}) \leq I_{0} \leq 1.1J(t_{0})\ \land \ 0 \leq \alpha_{1} \leq 1\ \land \ 0 \leq \alpha_{2} \leq \alpha_{1}\ \land \ 0.1(D-1)\Delta t \leq t_{{\rm c}}^{[\alpha]} \leq 0.9(D-1)\Delta t\ \land \ 0 \leq \beta \leq 1 \}$.
\end{itemize}

For the $\gamma$ problem,
\begin{itemize}
\item $P_{1}(\gamma)$ for the model without any change points is a constant in $\mbox{\boldmath{$\Theta$}}_{1} = \{\gamma |\ 0 \leq \gamma \leq 1 \}$
\item $P_{2}(\gamma_{1},\gamma_{2},t_{{\rm c}}^{[\gamma]})$ for the model with a change point is a constant in $\mbox{\boldmath{$\Theta$}}_{2} = \{\gamma_{1},\gamma_{2},t_{{\rm c}}^{[\gamma]} |\ 0 \leq \gamma_{1} \leq 1\ \land \ 0 \leq \gamma_{2} \leq \gamma_{1}\ \land \ 0.1(D-1)\Delta t \leq t_{{\rm c}}^{[\gamma]} \leq 0.9(D-1)\Delta t \}$.
\end{itemize}

The estimates of the parameters with a maximal likehood estimation are identical to those with a maximal a posteriori estimation. The number of random samples is $10^{5}$ for the marginalized likelihood selector.

\begin{figure}
\begin{center}
\includegraphics[scale=0.4,angle=-90]{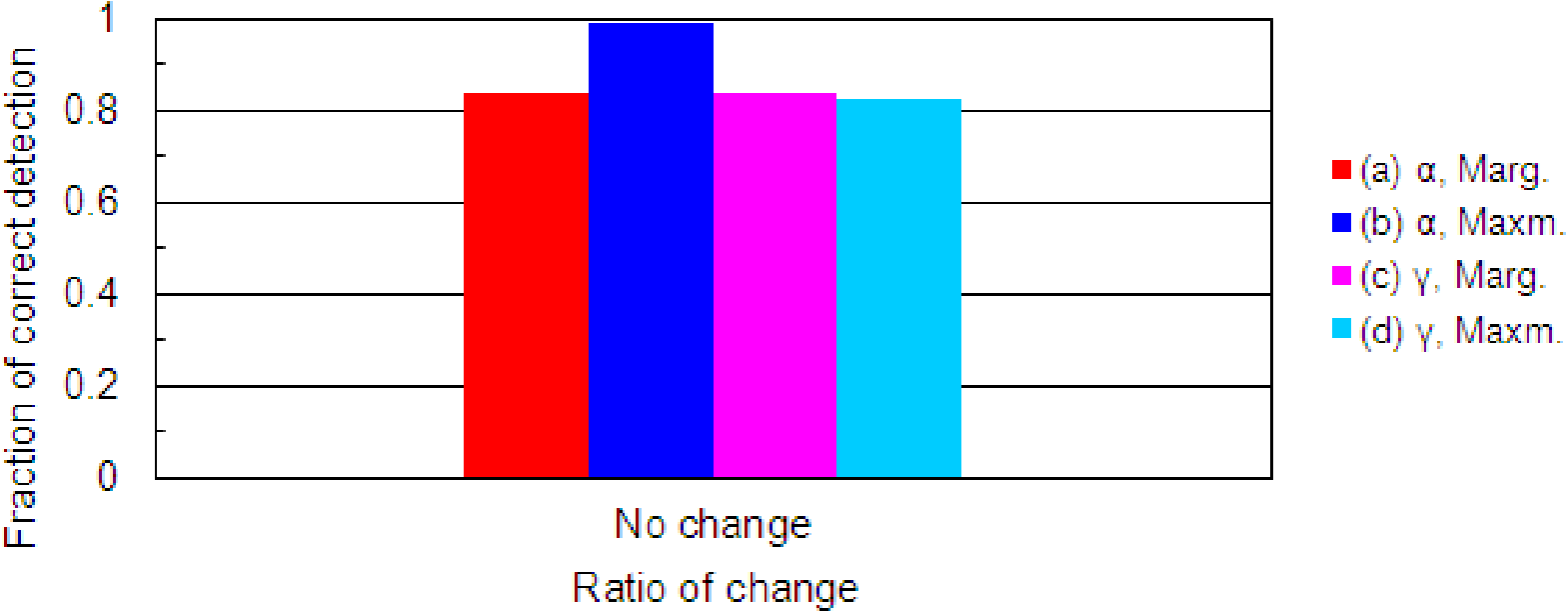}
\end{center}
\caption{Fraction of correct detection of the change in $\alpha$ and $\gamma$ from synthesized datasets of $D=30$ and $\Delta t=1$ for random digraphs of $N=30$ with the initial condition of $I_{0}(t_{0})=200$. The parameters $\alpha$ and $\gamma$ do not change at all. (a) Marginalized likelihood selector for the change in $\alpha$, (b) maximal likelihood selector for $\alpha$, (c) marginalized likelihood selector for $\gamma$, (d) maximal likelihood selector for $\gamma$.}
\label{201311011s}
\end{figure}

Figure \ref{201311011s} shows the fraction of correct detection of the change in $\alpha$ and $\gamma$ when $\alpha$ and $\gamma$ do not change at all. Correct detection means descriminating that the values of $\alpha$ and $\gamma$ do not change. The fraction is larger than 0.8. Both the model selectors work accurately.

\begin{figure}
\begin{center}
\includegraphics[scale=0.4,angle=-90]{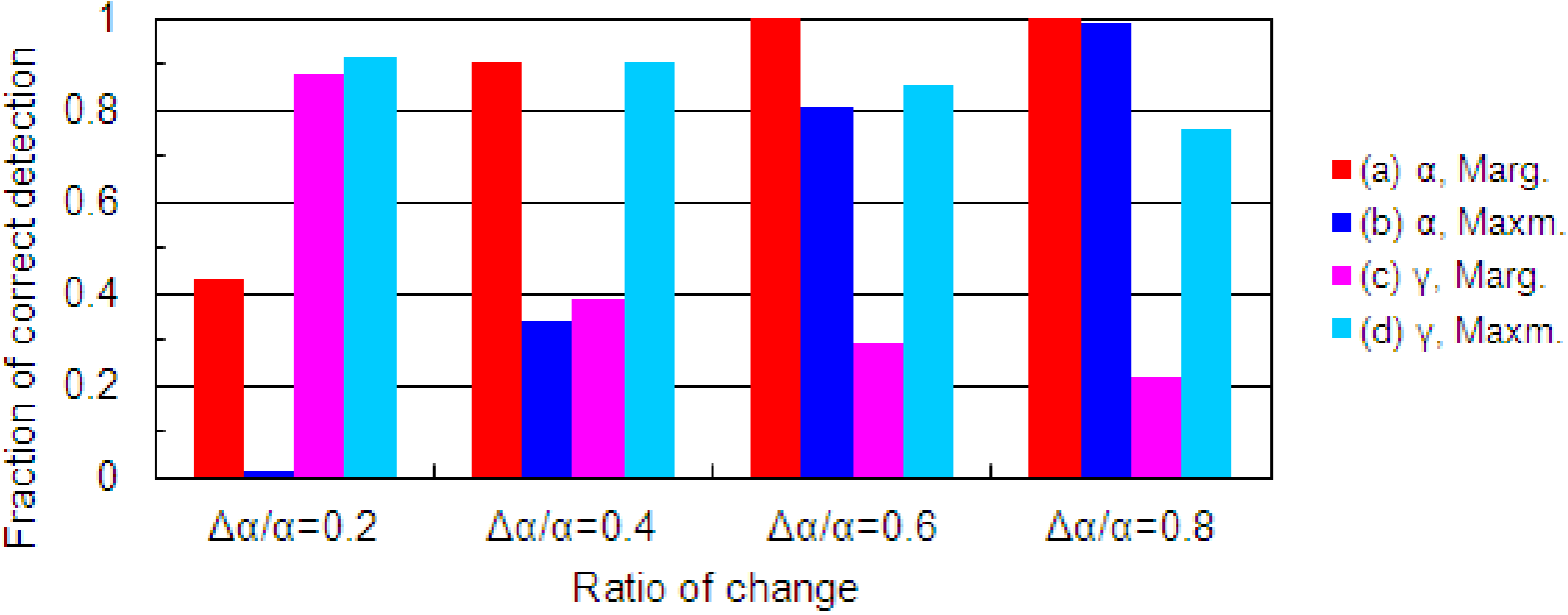}
\end{center}
\caption{Fraction of correct detection of the change in $\alpha$ and $\gamma$ from synthesized datasets of $D=30$ and $\Delta t=1$ for random digraphs of $N=30$ with the initial condition of $I_{0}(t_{0})=200$. The parameter $\alpha$ decreases from $\alpha_{1}=0.075$ to $\alpha_{2}=\alpha_{1}-\Delta \alpha$ at $t_{{\rm c}}^{[\alpha]}=0.5(D-1) \Delta t=14.5$, but $\beta=0.025$ and $\gamma=0.1$ do not change. The ratio of change $\Delta \alpha/\alpha_{1}$ is 0.2, 0.4, 0.6, or 0.8. (a) Marginalized likelihood selector for the change in $\alpha$, (b) maximal likelihood selector for $\alpha$, (c) marginalized likelihood selector for $\gamma$, (d) maximal likelihood selector for $\gamma$.}
\label{201311012s}
\end{figure}

\begin{figure}
\begin{center}
\includegraphics[scale=0.4,angle=-90]{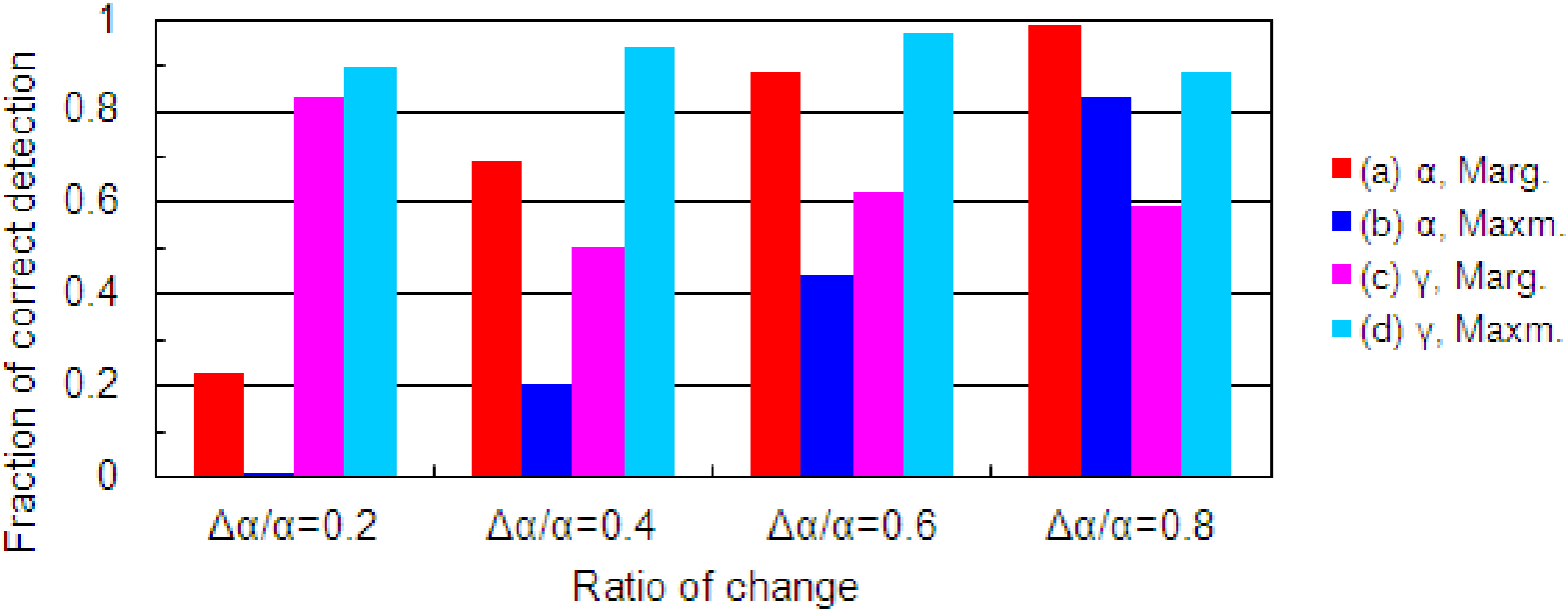}
\end{center}
\caption{Fraction of correct detection when $\alpha$ decreases at $t_{{\rm c}}^{[\alpha]}=0.2(D-1) \Delta t=5.8$. The other experimental conditions are the same as those for Fig.\ref{201311012s}.}
\label{201311013s}
\end{figure}

Figure \ref{201311012s} shows the fraction of correct detection of the change in $\alpha$ and $\gamma$ when $\alpha$ decreases at $t_{{\rm c}}^{[\alpha]}=0.5(D-1) \Delta t=14.5$. Correct detection means discriminating that the value of $\alpha$ changes but the value of $\gamma$ does not change. The fraction of correct detection of the change in $\alpha$ increases as the ratio of change increases. But the fraction of correct detection of the change in $\gamma$ decreases because the big change in the probability of infection is a noise component which destroys a sign of persons moving between sub-populations. In case of correct detection, $(\hat{\alpha}_{1}-\hat{\alpha}_{2})/\hat{\alpha}_{1}=0.67$ on the average with a standard deviation of $0.20$ when the true value is $\Delta \alpha/\alpha_{1}=0.8$, and $\hat{t}_{{\rm c}}^{[\alpha]}=14.8$ on the average with a standard deviation of $1.4$. The estimates of the parameters are is accurate. Figure \ref{201311013s} shows the fraction of correct detection of the change when $\alpha$ decreases at $t_{{\rm c}}^{[\alpha]}=0.2(D-1) \Delta t=5.8$. The fraction is close to that in Figure \ref{201311012s}. The location of a change point is not so critical to the model selectors. In case of correct detection, $(\hat{\alpha}_{1}-\hat{\alpha}_{2})/\hat{\alpha}_{1}=0.65$ on the average with a standard deviation of $0.14$ when the true value is $\Delta \alpha/\alpha_{1}=0.8$, and $\hat{t}_{{\rm c}}^{[\alpha]}=6.1$ on the average with a standard deviation of $1.1$. The estimates of the parameters are accurate too.

\begin{figure}
\begin{center}
\includegraphics[scale=0.4,angle=-90]{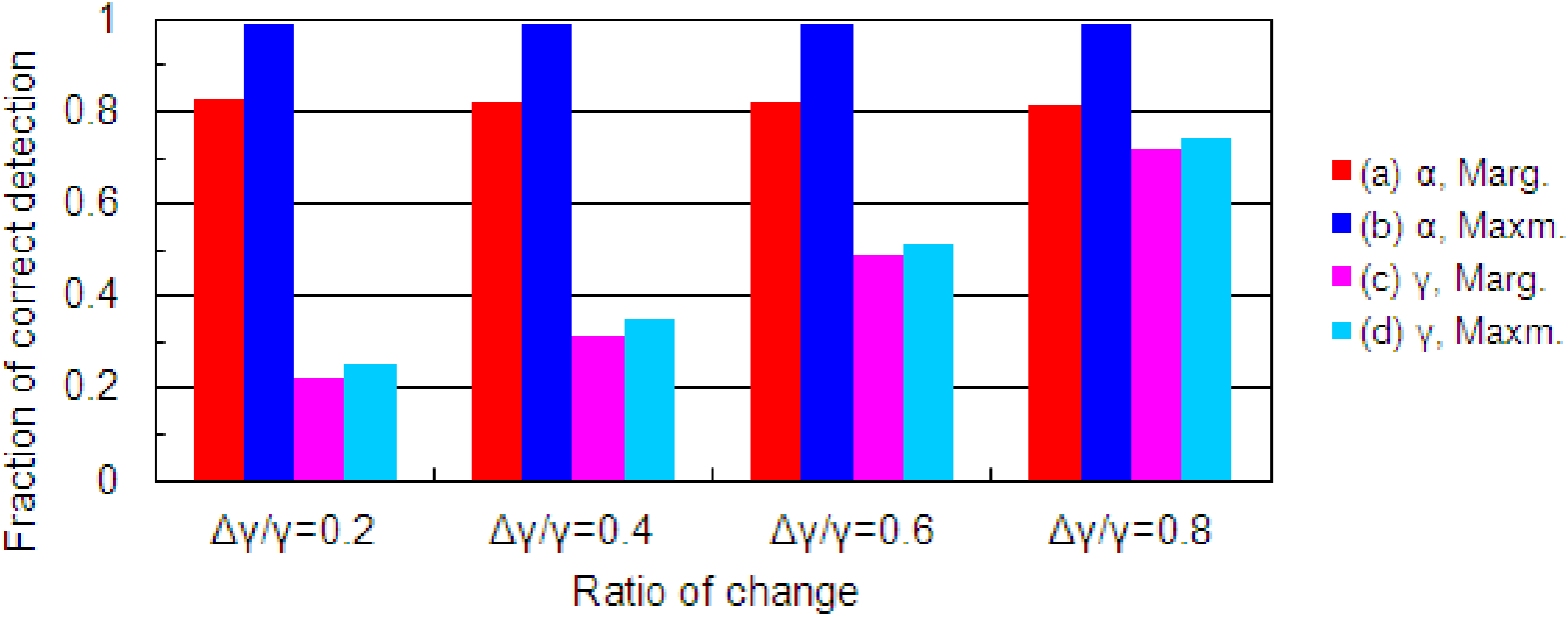}
\end{center}
\caption{Fraction of correct detection when $\gamma$ changes at $t_{{\rm c}}^{[\gamma]}=0.2(D-1) \Delta t$. The other experimental conditions are the same as those for Fig.\ref{201311012s}.}
\label{201311014s}
\end{figure}

\begin{figure}
\begin{center}
\includegraphics[scale=0.4,angle=-90]{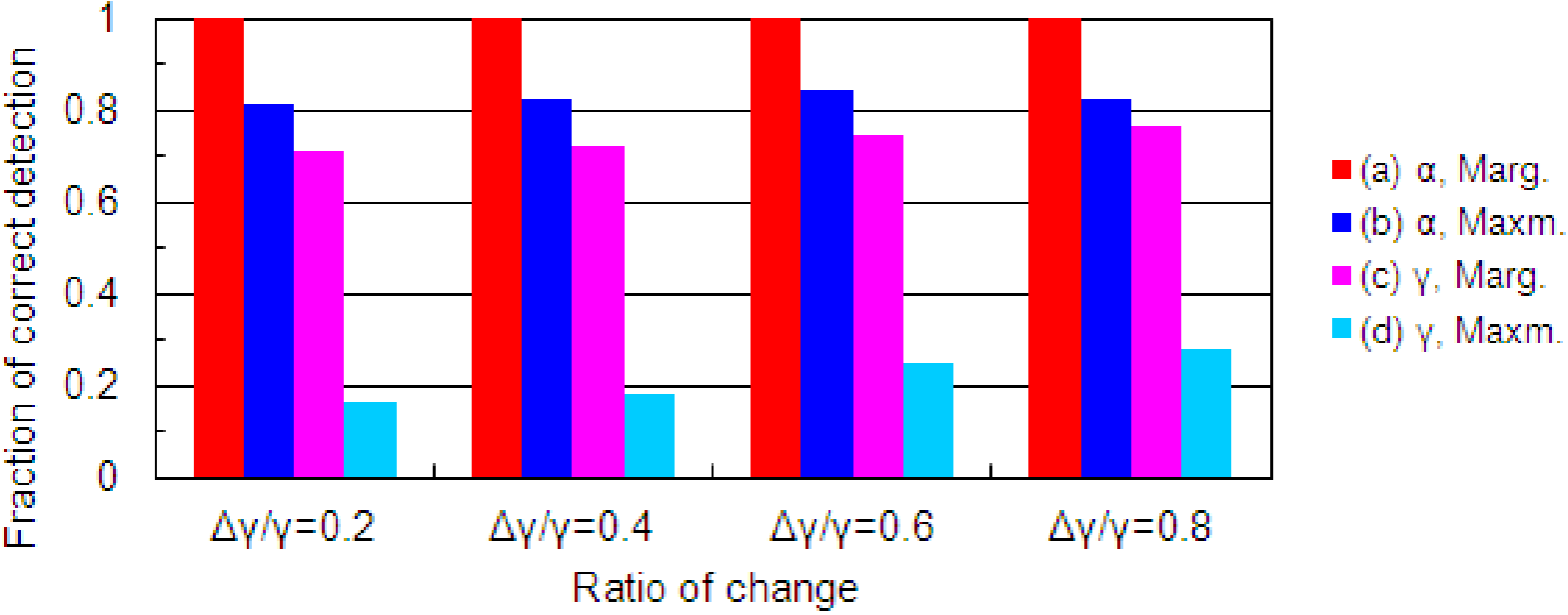}
\end{center}
\caption{Fraction of correct detection when $\gamma$ decreases at $t_{{\rm c}}^{[\gamma]}=0.2(D-1) \Delta t$, and $\alpha$ decreases by $\Delta \alpha/\alpha_{1}=0.6$ at $t_{{\rm c}}^{[\alpha]}=0.5(D-1) \Delta t$ as well. The other experimental conditions are the same as those for Fig.\ref{201311012s}.}
\label{201311015s}
\end{figure}

Figure \ref{201311014s} shows the fraction of correct detection of the change when $\gamma$ decreases at $t_{{\rm c}}^{[\gamma]}=5.8$. Correct detection means that discriminating $\alpha$ does not change but $\gamma$ changes. The fraction of correct detection of the change in $\gamma$ increases as the ratio of change increases. In case of correct detection, $(\hat{\gamma}_{1}-\hat{\gamma}_{2})/\hat{\gamma}_{1}=0.24$ on the average with a standard deviation of $0.088$ when the true value is $\Delta \gamma/\gamma_{1}=0.8$, and $\hat{t}_{{\rm c}}^{[\alpha]}=17.5$ on the average with a standard deviation of $7.0$. The estimates of the parameters is not so accurate as those for Figure \ref{201311012s}. This implies that it is difficult to solve the $\gamma$ problem accurately. The signal-to-noise ratio for the relative proportion of the number of cases in one sub-population to that in other sub-population is smaller than the ratio for the growing number of cases in the entire sub-populations. Figure \ref{201311015s} shows the fraction of correct detection of the change when $\gamma$ decreases at $t_{{\rm c}}^{[\gamma]}=5.8$ with a 60\% decrease of $\alpha$ at $t_{{\rm c}}^{[\alpha]}=14.5$. The marginalized likelihood estimator works more accurately than the maximal likelihood estimator when both the values of $\alpha$ and $\gamma$ change. The fraction is larger than 0.7.

\subsection{Real dataset}
\label{Realdataset}

Trend change is investigated in two real datasets with the uninformative prior probability density function in \ref{SynDatasets} and the marginalized likelihood selector with $10^{8}$ random samples. One dataset\footnote{World Health Organization, Cumulative number of reported probable cases of SARS, http://www.who.int/csr/sars/country/en/index.html (2003).} is the World Health Organization (WHO) archive on the SARS outbreak in 2003. The other dataset\footnote{World Health Organization, Situation updates - Pandemic (H1N1) 2009, http://www.who.int/csr/disease/swineflu/updates/en/index.html (2010).} is the WHO archive on the flu pandemic (H1N1 swine influenza A) in 2009. The dataset in the archives had been updated every day. It is a collection of time sequence data $J_{i}(t_{d})$ with $\Delta t=1$ day. The value of the elements of the adjacency matrix $\mbox{\boldmath{$l$}}$ is obtained from the time sequence of $\mbox{\boldmath{$I$}}(t_{d}) \approx \Delta \mbox{\boldmath{$J$}}(t_{d}) / \hat{\alpha} \Delta t$ with a maximal likelihood estimation\cite{Mae10}.

\subsubsection{SARS outbreak}

SARS is a respiratory disease in humans caused by the SARS corona-virus. The epidemic of SARS appears to have started in Guangdong of south China in November 2002. SARS spread from the Guangdong to Hong Kong in early 2003, and eventually nearly 40 countries around the world by July\cite{Lip03}. The WHO archives the cumulative number of reported probable cases. The target geographical regions in this study are those where five or more cases had been reported in a month since March 17. They are Canada, France, United Kingdom, Germany, Hong Kong, Malaysia, Taiwan, Singapore, Thailand, United States, and Vietnam. The number of geographical regions is $N=11$. The number of data is $D=31$.

The change in $\gamma$ is detected in the dataset. Selecting this model is decisive. The estimated value decreases by 33\% at the estimated change point $\hat{t}_{{\rm c}}^{[\gamma]}=3.2$ which is March 20. The change point was one week after the WHO worldwide alert on March 12. This implies that potential travelers refrained from international travels in the middle of March, and that the alert was effective in reducing the movements of infectious persons across borders and controlling cross-border exposure.

The change in $\alpha$ is detected too. Selecting this model is substantial. The estimated value decreases by 29\% at the estimated change point $\hat{t}_{{\rm c}}^{[\alpha]}=16.0$ which is April 2. The consequent reproductive ratio decreases from $\hat{R}_{1}=1.5$ to $\hat{R}_{2}=1.0$. According to the field-based medical case studies, $R$ went down to 1 in late March in Hong Kong\cite{Ril03}. This result is in good agreement with the estimated change point on April 2. This implies that the WHO worldwide alert barely affected local outbreaks directly for three weeks. Public health intervention took few immediate effects after the WHO worldwide alert. This time gap would have been shorter if the standards for diagnosis and treatment had been established more quickly and the public health authorities had controlled local exposure more successfully.

\subsubsection{Flu pandemic}

The flu pandemic was a global outbreak of a new strain of the H1N1 swine influenza A virus. The virus appeared in Veracruz in southeast Mexico in April 2009. The pandemic spread to United States and Canada immediately, and then to the South American countries, West European countries, and Pacific Rim countries. It began to decline in November. The WHO archives the cumulative number of the reported laboratory-confirmed cases. The target geographical regions in this study are those where five or more cases had been reported in about three weeks since April 28. They are Australia, Belgium, Brazil, Canada, Chile, China, Colombia, Costa Rica, Ecuador, El Salvador, France, Germany, Israel, Italy, Japan, Mexico, New Zealand, Panama, Peru, Spain, United Kingdom, and United States. The number of geographical regions is $N=22$. The number of data is $D=25$. The dataset is smoothed with a moving average filter whose window size is $W=3$.

Any changes in $\gamma$ are not detected in the dataset. International travels and cross-border exposure were not affected although the WHO raised the global pandemic alert level to the phase 5 on April 29, which signals that community transmission is sustained across national borders. This is confirmed by the estimate by the United Nations World Tourism Organization that the tourism declined by only 4 \% in Mexico\footnote{World Health Organization, Public health measures during the influenza A (H1N1) 2009 pandemic, http://www.who.int/influenza/resources/documents/health\_mesures\_h1n1\_2009/en/index.html (2010).}. The public awareness of the pandemic might not be as acute as that at the time of the SARS outbreak probably because the flu in 2009 was mild in contrast to the severe flu in 1918, 1957, and 1968.

But the change in $\alpha$ is detected.  Selecting this model is decisive. The estimated value decreases by 52\% at the estimated change point $\hat{t}_{{\rm c}}^{[\alpha]}=7.5$ which is May 5. The consequent reproductive ratio decreases from $\hat{R}_{1}=3.2$ to $\hat{R}_{2}=1.5$. ${R}_{2}$ is nearly the same as the basic reproductive ratio obtained from epidemiological and genetic analyses in Mexico\cite{Fra09} and the United States\cite{Jhu11}. Local public health authorities in many countries started mandatory school closures, requested cancellation of large mass gatherings, and took other possible social distancing measures in May\cite{Cho11}. The implemented intervention was effective in controlling local exposure.

\section{Discussion}

Although the findings for the SARS outbreak and flu pandemic are still merely narrative evidence, the change points are indicative of the potential impact of public health intervention on cross-border epidemic transmission. The cross-border epidemic transmission ensues from a generally complex interplay between movements and infection. These stochastic processes may have adverse effects. Cross-border movements prompt the geographical distribution of cases to reach equilibrium while infection fuels local ourbreak. In this study, the complexity is resolved by decomposing the Langevin equations in eq.(\ref{apdI/dt}) into smaller dimensional model selection sub-problems so that they can be solved individually and sequentially with the state-of-the-art model selectors in Bayesian statistics. {\bf The cross-border movements and local outbreak of a pathogen are a special case of the diffusion and  reaction of a substance. The method is applicable potentially to solving an inverse problem and  contributing to significant findings for a dynamic system, whose time evolution as a stochastic reaction-diffusion process is formulated by Langevin equations.}

Public health authorities need to understand the efficacy of raising public awareness, social distancing, and other measures when they design an effective public health program and make an urgent decision on the verge of massive community transmission. Raising public awareness includes cancellation or postponement of travels, pre-travel health advisories, giving advices via mass media on hand washing, personal hygiene, cough etiquette, the use of face masks, hand rubs, and vaccine. In addition, border health screening, arrival and departure monitoring, dedicated ambulances, isolated hospital wards, and disinfection on public transport can be implemented. It is anticipated in the current practices that quarantine on board and temperature screening at airports will stop infectious travelers from entering across national or regional borders. Such a naive anticipation gives rise to much controversy about the economic efficiency. The anticipation is tested with observations when the intervention is implemented actually. The efficacy is quantified by the change in the probability parameters, and the economic efficiency is calculated. The public health authorities can accumulate ground information from such testing to organize an internationally shared extensive knowledge base on miscellaneous individual or combined public health intervention, its rational and anticipated outcome, empirical evidence on its efficacy, and possible reasons for the gap between the anticipation and observations, along with the supplementary knowledge from field-based medical case studies, epidemiological, and genetic studies under many demographic circumstances.

In obtaining more minute ground information for the knowledge base, the method in this study will be extended to more complicated mathematical models so that multiple change points can be detected, localized change points can be analyzed, and multiple geographic resolutions can be incorporated. The combined public health intervention may take effects mutiple times on different datas. The trend change in the entire population may arise from large localized trend changes in only a few sub-populations (an anomaly which violates the empirical law in eq.(\ref{gammacalc})). Adjusting resolutions may be essential in the drilling-down from country level course analysis to province, city, and district level fine analyses. These are the topics for future studies.

\end{document}